# Electron-Phonon Mediated Superconductivity in 1*T*−*MoS*$_2$ and Effect of Pressure on the Same

Jagdish Kumar[1] & Harkirat Singh[2]

[1]Department of Physics & Astronomical Science, Central University of Himachal Pradesh, Shahpur, Kangra, H.P. 176206 ( India )

[2]National Institute of Tecnology, Srinagar, Hazaratbal, Kashmir-190006 ( India )

E-mail: jagdishphysicist@gmail.com, harkirat@nitsri.ac.in

Abstract. Density functional theory (DFT) based ab-initio calculations of electronic, phononic and superconducting properties of 1*T* −*MoS*$_2$ are reported. The phonon dispersions are computed within density-functional-perturbation-theory (DFPT). We have also computed Eliashberg function $\alpha^2F(\omega)$ and electron-phonon coupling constant $\lambda$ from the same. The superconducting transition temperature ($T_c$) computed within McMillan-Allen-Dynes formula is found in good agreement with recent experimental report. We have also evaluated effect of pressure on the superconducting behaviour of this system. Our results show that 1*T* − *MoS*$_2$ exhibit electron-phonon mediated superconductivity and the superconducting transition temperature rises slightly with pressure and then decrease with further increase in pressure.

1. Introduction

Since its discovery in 1911, superconductivity has been a mysterious phenomenon [1]. It took almost half a century to establish first systematic theoretical model of electron-phonon-electron coupling to explain superconductivity [2]. A lot has been done since then in the field of superconductivity and many unconventional discoveries have been reported in the field [3] , [4], [5], [6], [7], [8], [9]. These rigorous attempts reveals two major concerns in the field of superconductivity. First is to invent superconductors having transition temperature close to room temperature and second to investigate possible mechanism of superconductivity beyond electron-phonon pairing. Till few years back, it was believed that electron-phonon based superconductivity cannot be observed at high temperature [10] and superconductors with exotic pairing mechanisms are the key to invent near room temperature superconductivity [11]. However, N. W. Ashcroft had been speculating higher superconducting transition temperatures ($T_c$) in hydrogen rich materials since long [12], [13]. Many theoretical studies have been done in this direction and $T_c$ as high as 200K has been predicted in such systems [14], [15], [16]. The scenario became really interesting when superconductivity was experimentally discovered at 203K in $H_3S$ [17]. Since then, many theoretical studies have predicted high $T_c$ in other hydride materials as well [18], [19], [20] and experimental verification have also been reported for some of these systems [21], [22]. Recent observation of room temperature superconductivity in carbonaceous sulphur hydride around 267 GPa has boosted interest in the field of superconductivity [23]. This has opened up whole new possibilities of predicting high temperature superconducting materials. However, the structures of all studied hydride materials are stable at very high pressures (hundreds of GPa) [24].

Recently, superconductivity has been reported experimentally in $1T − MoS_2$ which is stable at ambient pressure [25]. The transition temperature ($T_c$) is although found to be 4.2K which is very low in comparison to recent hydride systems. To the best of our knowledge, there has been no theoretical study to investigate mechanism of superconductivity in $1T −MoS_2$. In this paper we have investigated electronic and phononic properties of this system and also computed superconducting transition temperature within McMillan-Allen Dynes approach. Our calculations have found the transition temperature of 3.67K which is in reasonable agreement with experimental report. This indicates that superconductivity in this system is electron-phonon mediated.

Motivated by the fact that higher transition temperature is observed in hydride superconductors at high pressure, we have investigated pressure dependence of superconducting transition temperature for $1T −MoS_2$. Our calculations indicate that transition temperature shows slight rise for a pressure of 2 GPa but decrease with further increase in pressure.

2. Structural and Computational Details

It has been reported that $1T −MoS_2$ crystallize in trigonal space group (No. 164) P-3m1 with lattice parameters a=b=3.190 (6.0279 bohr), c=5.945 (11.2339 bohr) [25]. The Mo atoms occupy 1a position with coordinates (0.0, 0.0, 0.0) and S atoms occupy 2d position having coordinates (1/3, 2/3, z). The experimental value of z is reported to be 0.2488(7) in recent experiment on single crystals [25]. We have done total energy calculations to obtain energy at different values of volume and obtained the calculated value of structural parameters (refer to Table 1).

All the calculations have been done within density functional theory as implemented in ELK code [26]. The effect of periodic potential on electrons is considered within FullPotential Linearized Augmented Plane Wave (FP-LAPW) approach [27]. The electron-electron exchange and correlations have been incorporated using local density approximation (LDA) as implemented by Perdew and Wang using Ceperlay-Alder formalism [28]. A fine grid of 16 × 16 × 8 containing 213 k-points has been taken in irreducible Brillouin zone (IBZ) for doing ground state total energy calculations. The same set of k-points was used for obtaining minimum force position of atoms. We have used maximum length of G vector for expanding interstitial density $G_{max}$ = 20, $R_{MT} \times |G+k|$ = 7.0. Muffin tin radius of 2.3 a.u., and 2.0 a.u. were used for Mo and S atoms respectively. The convergence with respect

Table 1. Computed structural parameters at different pressure.

| | | 0 GPa | 2 GPa | 4 GPa | 10 GPa | 25 GPa |
|---|---|---|---|---|---|---|
| Lattice parameters (in bohr) | a | 5.8933 | 5.8784 | 5.8693 | 5.8533 | 5.8332 |
| | c | 11.4919 | 11.4629 | 11.4452 | 11.4139 | 11.3747 |
| $z_{chalcogen}$ | | 0.26022 | 0.26062 | 0.26145 | 0.26361 | 0.26520 |
| $\lambda_{el-ph}$ | | 0.6578 | 0.6688 | 0.6627 | 0.6333 | 0.5750 |
| $\omega_{log}(Hartree)$ | | 6.136$10^{-4}$ | 6.121$10^{-4}$ | 6.292$10^{-4}$ | 6.707$10^{-4}$ | 7.202$10^{-4}$ |
| $\Omega_{rms}$ | | 8.822$10^{-4}$ | 8.841$10^{-4}$ | 8.936$10^{-4}$ | 9.201$10^{-4}$ | 9.528$10^{-4}$ |
| $T_c$ McMillan-Allen-Dynes (K) | $\mu^*$ = 0.15 | 3.50 | 3.71 | 3.69 | 3.31 | 2.36 |
| | $\mu^*$ = 0.138 | 4.00 | 4.22 | 4.21 | 3.85 | 2.87 |
| $\Theta_D(K)$ | | 600 K | | | | |
| $N(E_F)(States/Hartree-u.c.)$ | | 34.06 | 33.50 | 33.04 | 31.71 | 30.06 |
| $\gamma_b(mJ/mole-K^2)$ | | 2.95 | 2.90 | 2.86 | 2.75 | 2.60 |
| $\gamma = (1+\lambda)\gamma_b$ | | 4.89 | 4.84 | 4.75 | 4.49 | 4.09 |

to various computational parameters was ensured.

The atomic positions were relaxed to a force tolerance of $1.5 \times 10^{-7} eV/\text{Å}$ before computing phonon dispersions and other related parameters. For computing phonon spectrum, Eliashberg function $\alpha^2 F(\omega)$, electron-phonon coupling constant , we have used a Q-point mesh of 4×4×2 containing 8-Q points in IBZ. From the parameters obtained, we have computed superconducting transition temperature using McMillan-Alen-Dynes formula [29]. For computing electron-phonon matrix elements and hence coupling constant, we have used very dense grid of 32×32×16 k-points yielding to 1513 k-points IBZ. The pressure dependent electronic properties and superconducting transition temperature are calculated by employing lattice parameters at high pressure which are obtained by fitting energy versus volume data to universal equation of states [30] (see section 4.1 for details).

3. 3. Electron-phonon interaction and superconductivity within DFT

*3.1. Density Functional Perturbation Theory: A quick overview*

The state of the art density functional theory can be employed to study not only ground state properties of a material but also the phonon spectrum and electron-phonon coupling. DFPT is a very efficient approach to compute phonon dispersions and electron-phonon coupling matrix elements within density functional theory [31], [32], [33]. To compute lattice dynamics of a crystalline system within DFPT, one start by relaxing the crystal structure to lowest energy (say $E_{tot}^0$) state by minimizing all the forces on the atoms. Then small displacements, $u_{\alpha i}(R)$ and $u_{\beta j}(R')$ of atoms (marked by α and β) having relaxed positions $R$ and $R'$, are generated corresponding to polarisations $i$ and $j$. Then one can compute corresponding self consistent energy of crystal with displaced atoms by employing Kohn-Sham density functional theory.

The energy of the crystal with atom displaced from their equilibrium position can be written as:

$$E_{tot}^{disp} = E_{tot}^{(0)} + \frac{1}{2}\sum_{\substack{\alpha,i,R \\ \beta,j,R'}} C_{\alpha i,\beta j}(R - R')\, u_{\alpha i}(R) u_{\beta j}(R') + \cdots \quad (1)$$

If the displacements are small in magnitude, one can restrict to harmonic form of above expression and hence neglect higher order terms. By employing energies computed from density functional theory, one can obtain the coefficients $C_{\alpha i,\beta j}(R - R')$ of equation (1) as given below:

$$C_{\alpha i,\beta j}(R - R') = \frac{\partial^2 E}{\partial u_{\alpha i}(R)\, \partial u_{\beta j}(R')} \quad (2)$$

These real space coefficients of equation (2) can be used to obtain the dynamical matrix corresponding to wave vector (q) of phonon perturbation as follows:

$$D_{\alpha i,\beta j}(q) = \frac{1}{\sqrt{M_\alpha M_\beta}} \sum_R e^{iq\cdot R} C_{\alpha i,\beta j}(R) \quad (3)$$

$M_\alpha$ and $M_\beta$ are masses of atoms α and β respectively. One can then write Eigen value equation for a particular phonon wave vector as below:

$$D_{\alpha i,\beta j}(q) e_{\alpha i}(q) = \omega_i^2(q) e_{\alpha i}(q) \quad (4)$$

The equation (4) results to phonon frequency $\omega_i$ for that particular polarisation mode at phonon wave vector q-vector. The Eigen vectors obtained above can be used to compute electron-phonon coupling matrix elements as follows:

$$g_{nik+q,jk} = \frac{1}{\sqrt{2M_\alpha \omega_n(q)}} \langle \phi_{i,k+q} | e_n(q) \cdot \Delta V_{KS}(q) | \phi_{j,k} \rangle \quad (5)$$

where $\Delta V_{KS}(q)$ is effective potential derivative and is given by [34]:

$$\Delta V_{KS}(q) = l_{qi} \sum_{\alpha} \sqrt{\frac{M_0}{M_\alpha}} e_{\alpha^i}(q) \partial_{\alpha i,q} V_{KS} \quad (6)$$

Where $l_{qi}$ is called zero-point displacement as is given by $l_{qi} = \sqrt{\frac{\hbar}{2M_0 \omega_{qi}}}$. $M_0$ is an arbitrary reference mass and is usually taken to be mass of proton. $\partial_{\alpha i,q} V_{KS}$ follows from Kohn-Sham potential $v_{KS}(= v^{ne} + v^H + v^{xc})$ as below:

$$\partial_{\alpha i,q} V_{KS} = \sum_p e^{-iq(r-R_p)} \left. \frac{\partial v_{KS}}{\partial u_{\alpha i}(R)} \right|_{r-R_p} \quad (7)$$

The summation in equation (7) runs over all the p unit cells in crystal as per Born-Von Karman boundary conditions.

Once the electron-phonon coupling matrices are defined, one can then compute phonon linewidths $\gamma_n(q)$ as given below:

$$\gamma_n(q) = 2\pi \omega_n(q) |g_{ik+q,jk}^n|^2 \delta(\varepsilon_{i,k+q} - \varepsilon_F) \delta(\varepsilon_{j,k} - \varepsilon_F) \quad (8)$$

From which one can obtain Eliashberg function as given below:

$$\alpha^2 F(\omega) = \frac{1}{2N(\varepsilon_F)} \sum_{q,n} \frac{\gamma_n(q)}{\omega_n(q)} \delta(\omega - \omega_n(q)) \quad (9)$$

$N(\varepsilon_F)$ is electronic density of states at Fermi level, $\alpha(\omega)$ is called electron-phonon coupling strength and $F(\omega)$ is phonon density of states. Once we have Eliashberg function, we can obtain electron-phonon coupling constant $\lambda$ as follows:

$$\lambda = 2 \int_0^\infty d\omega \frac{\alpha^2 F(\omega)}{\omega} \quad (10)$$

The above described formalism of DFPT is implemented in Elk code which has been employed for obtaining results presented in this paper.

*3.2. McMillan-Allen-Dynes formula for superconducting transition temperature*

The McMillan-Allen-Dynes formula provides excellent approximation for computing superconducting transition temperature of a material exhibiting phonon mediated superconductivity [29]. According to this formulism the superconducting transition temperature is given by:

$$T_c = \frac{\omega_{log}}{1.2 k_B} \exp\left(\frac{-1.04(1+\lambda)}{\lambda - \mu^*(1+0.62\lambda)}\right) \quad (11)$$

$\lambda$ can be obtained by employing DFPT as given by equation (10). $\omega_{log}$ is logarithmic averaged phonon frequency and is given by:

$$\omega_{log} = exp(\frac{2}{\lambda} \int d\omega \frac{\alpha^2 F(\omega)}{\omega} ln(\omega)) \tag{12}$$

This can also be obtained by employing $\alpha^2F(\omega)$ obtained from DFPT as given in equation 9. $\mu^*$ is called Morel-Anderson pseudopotential and is screened coulomb potential of the electrons. The empirical value of $\mu^*$ is found around 0.15 for most of the conventional superconductors.

## 4. Results and Discussion

### 4.1. Structural Properties

The crystal structure of $MoS_2$ has two independent parameters *a* and *c*. To obtain values of these parameters in ab-initio manner we have computed total energy for various volumes around experimentally reported value. For each value of volume (say V), we have systematically varied c/a and computed total energy using DFT based calculations. Afterwards, value of c/a for lowest energy was determined for respective volume. The same process was repeated for different volumes and an overall minimum was obtained for energy versus volume as shown in Fig. 1. As depicted in plots, the volume corresponding to lowest energy is 51.23 $A°^3$ (345.672$bohr^3$) with corresponding value of c/a equal to 1.95. The computed value of c/a is in reasonably good agreement with experimental value of 1.86 [25].

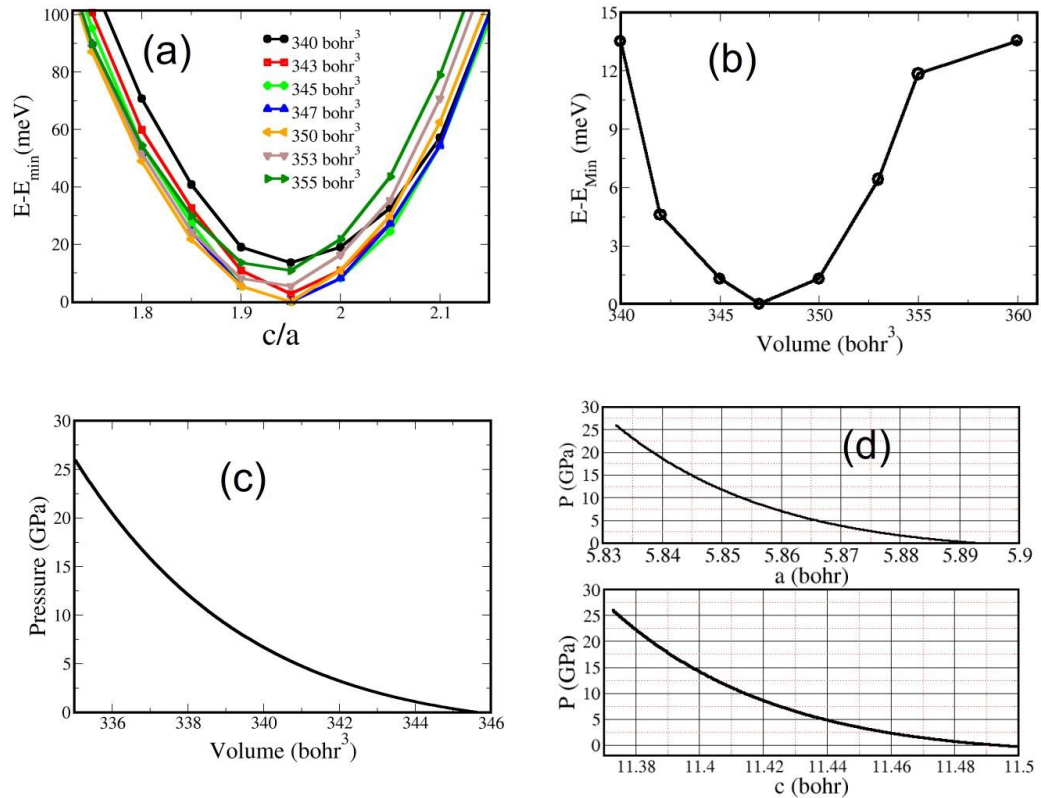

Figure 1. (a) Total energy versus c/a plotted for different volumes for 1T −MoS$_2$. (b) The energy corresponding to overall minima of each volume is plotted at function of volume. This energy versus volume data is used to fit equation of state The corresponding value of c/a is 1.95.

From the computed values of volume and c/a, the value of parameter a and c was computed using relations $a = (\frac{V}{\frac{c}{a}sin(\pi/3)})^{1/3}$ and $c = a(c/a)$.

The obtained values of a and c are 3.119 $A°$ (5.8938 bohr) and 6.082 $A°$ (11.4928 bohr) respectively, are in reasonably good agreement with experimental values of 3.19 $A°$ and 5.945 $A°$ [25]. The computed unit cell volume of 51.23 $A°^3$ is again very close to experimental value of 52.39 $A°^3$. The small deviation from experimental values can be related to intrinsic inability of local density approximation to produce exactly accurate structural parameters. We also found that there is no significant change in value of c/a as the volume change. The overall lowest energy for each volume is obtained and the energy versus volume data is shown in Figure 1(b). This data has been used further to fit universal equation of state [31], as implemented in ELK code [26]. The equation of states (EOS) can give various structural parameters such as lowest energy volume ($V_0$), bulk modulus ($B_0$) and pressure dependent volume, energy and enthalpy. The value of $V_0$ obtained is 51.23 $A°^3$ and bulk modulus was found to be 180.98 GPa. The interpolated values of pressure versus volume as obtained from fitted equation are plotted in Figure 1 (c). The effect of pressure on lattice parameters a and c is also shown in Figure 1 (d). The parameters obtained after fitting this equation are then used to interpolate structural parameters of 1T −MoS$_2$ at high pressures.

4.2. 1T-MoS2 at ambient pressure

The electronic band structure and density of states for 1T −MoS$_2$ is shown in Figure 2. The Mo-d and S-p states are distributed over broad energy range around Fermi level. This indicates strong hybridization among Mo-d and S-p states. It can also be seen that the density of states near Fermi level have contribution from both Mo-d and S-p character. Another key feature observed near Fermi level is presence of electron type bands near A and Γ point. At A high symmetry point there are two types of electron pockets having different curvature and hence effective mass. These features indicate that 1T − MoS$_2$ should have electrons as dominant charge carriers. We also observed that there is a peak in electronic density of state curve at energy slightly higher than Fermi level. This indicates that in case superconductivity in this compound is mediated within BCS mechanism then small electron doping would increase density of states at Fermi level and hence superconducting transition temperature should be increased.

The phonon dispersions and density of states are given in Figure 3. The first feature to notice is that there are no negative frequencies in phonon dispersion plots which indicate that structure is stable. We found that three acoustical branches namely longitudinal acoustical (LA), transverse acoustical (TA and T$A'$) branches can have highest energies of up to 30 meV. Near Γ point the all the phonon branches show linear dispersions representing their classical limits. The TA and T$A'$ branches are degenerate near low wavevector limits and have higher slope than LA branch. The highest energy, $\hbar\omega$, for LA branch is around 17 meV which results in Debye temperature of 200 K (using $T_D = \hbar\omega/k_B$) for LA branch. At high symmetry point A, the TA and T$A'$ branches lose their degeneracy and disperse differently along say A-K and A-H directions. The highest phonon energy of 30meV indicates Debye temperature of 350 K for TA branches. There is a peak in phonon density of states around 15meV region. This peak may be attributed to sudden lowering of energy states for TA branch around ΓM and M-A regions. Referring to optical branches, we found that there is slight overlap in energy states of optical and acoustical branch around Γ-M and L-A region. The highest frequency of the phonons in this compound is around 52 meV which corresponds to Debye temperature of 600K. This value of upper cut off energy of optical phonons for 1T −MoS$_2$ is although very small in comparison to recently discovered high pressure hydrides which can have phonon energies around 200meV (for *H$_3$S*) and hence possess very high transition temperature of around 204K. This indicates that $1T-MoS_2$ should have relatively very low transition temperature which is in agreement with recent experimental report [25]. Also there are sharp peaks in density of states for optical phonon branches. Such high density of states for optical phonons indicates possibility of stronger electron phonon coupling and hence superconductivity.

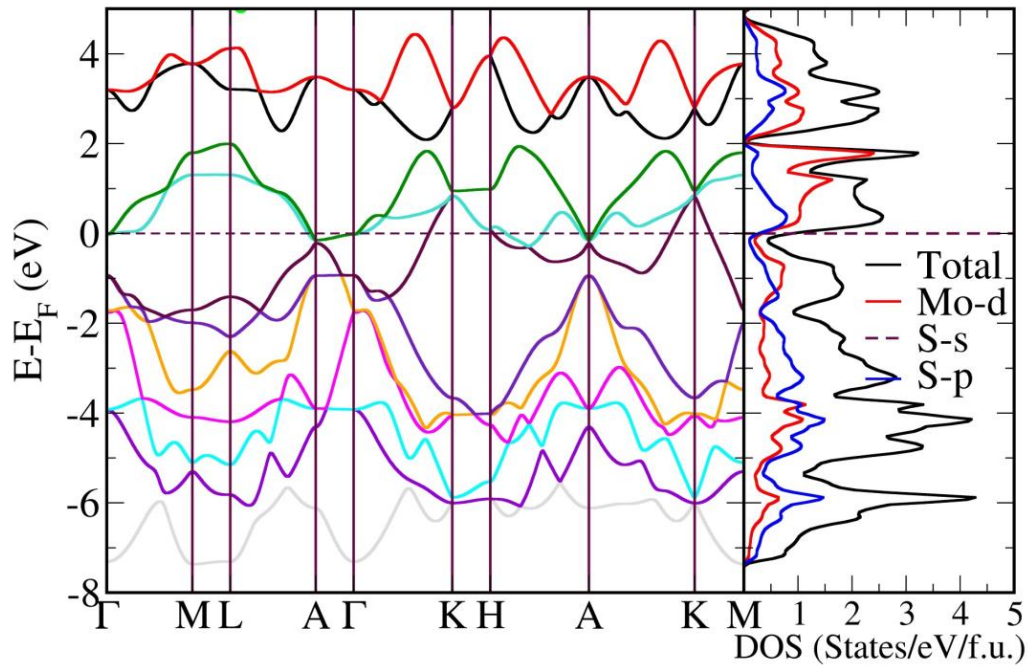

Figure 2. Electronic band structure and density of states for 1T-MoS2 at ambient pressure (0 GPa)

From phonon density of states, $D(\omega)$, we can also compute thermal energy of a solid due to phonons as given below:  $U = \sum_p \int d\omega D_p(\omega) \frac{\hbar\omega}{\left(e^{\hbar\omega/kT}-1\right)}$ (13)

The summation runs over all the polarization modes. The phononic contribution to specific heat then follows by taking derivative of equation 13 with respect to temperature ($C_{ph} = dU/dT$).

We have computed phononic specific heat as function of temperature for $1T-MoS_2$ by employing an inbuilt algorithm in Elk code [26]. The plot of specific heat is shown in Figure 4. It can be seen from this figure that above Debye temperature of 600K, the phononic specific heat approaches the value of 72 $mJmol^{-1}K^{-1}$. This value turns out to be close to

DulongPetit limit of 25 $mJmol^{-1}K^{-1}$. In *MoS₂* unit cell, total atoms are three and hence to obtain specific heat per mole atoms we have to divide by 72 Joule/mole/K number of atoms giving 24 $mJmol^{-1}K^{-1}$. Our calculated value of phononic contribution to specific heat is also in good agreement with earlier calculations [35].

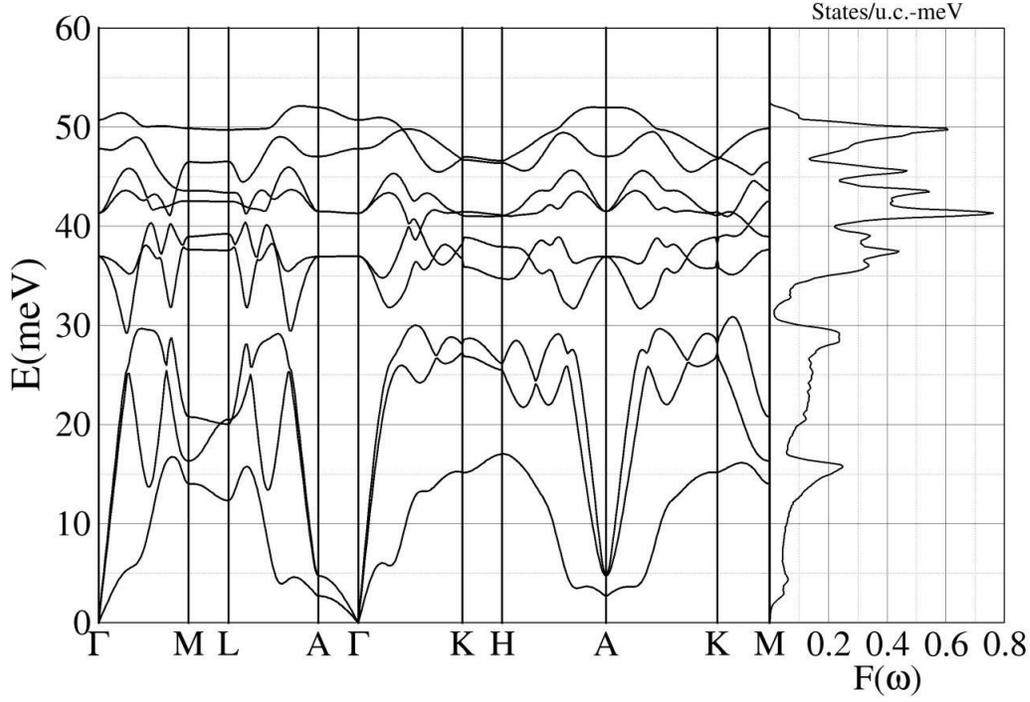

Figure 3. Phonon dispersions for 1T MoS2 at ambient pressure (Left) and phonon density of states *F(ω)* ( right ).

We can also obtain electronic contribution to specific heat by employing Somerfield approximation given by:

$$C_{el} = \frac{2}{3}\pi^2 N(E_F) k_B^2 T = \gamma_b T \qquad (14)$$

where *N(E_F)* is electronic density of states at Fermi level, $k_B$ is Boltzmann constant and $\gamma_b$ is also known as Somerfield constant as it can be estimated from *N(E_F)* band structure calculations. The above approximation of electronic specific heat modifies due to electronphonon interaction as $C_{el}$ =$\gamma_b$(1+λ)T =γT, where λ is electron-phonon coupling constant. We have computed both γ and $\gamma_b$ using *N(E_F)* and λ (details described in next section) values and are summarised in Table 1. The Somerfield contribution to electronic specific heat is also plotted in Figure 4.

At low temperatures the normal state specific heat of a superconductor can be fitted to simple equation:

$$C_{tot} = C_{ph} + C_{el} = \beta T^3 + \gamma T \qquad (15)$$

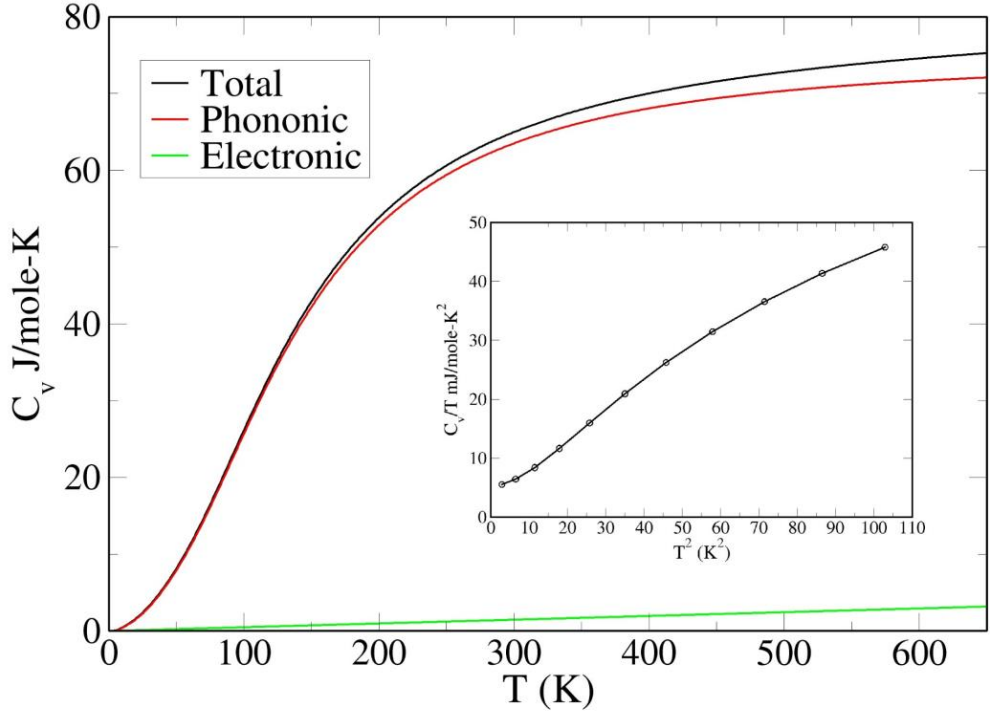

Figure 4. Specific heat of 1T −MoS₂ as function of temperature. Red curve shows phonon contribution to specific heat as computed using Elk code, green curve is obtained using Somerfield approximation as described in text, black curve shows sum of the two and represent total specific heat.

The first term on right hand side of equation 15 is Debye low temperature approximation to phononic specific heat and second term is Somerfield term for electronic specific heat. The inset of Figure 4 shows plot of $C_{tot}/T$ versus $T^2$ and the plot is almost straight line. However, the slope $\beta$ and intercept $\gamma$ obtained from our calculations is significantly different than those reported in experimental data reported by Fang et al. [25]. Their measurements show an intercept of 1.0 *mJmol⁻¹K⁻²* whereas intercept $\gamma$ from our calculations is 5.0 *mJmol⁻¹K⁻²*. Similarly, the value of slope from experimental data of Fang et al is 0.01 *mJmol⁻¹K⁻⁴*. This value is way different from the value 0.4 *mJmol⁻¹K⁻⁴* what our calculations results. We could not find more reports in literature on specific heat data to cross check our values. The convergence of high temperature value of specific heat obtained from our calculations to Dulong-Petit limit seems to validate our calculations. Therefore, more studies on specific heat of 1T −MoS$_2$ are required to shed more light on this ambiguity.

Figure 5 shows the plots of electron phonon coupling strength $\alpha(\omega)$, Eliashberg function $\alpha^2 F(\omega)$ and phonon density of states computed for 1T −MoS$_2$ at various pressures (to be discussed in next section) by employing the approach of DFPT. It can be seen that there is a sharp peak in $\alpha^2F(\omega)$ between energy range of 10-20 meV which lies in acoustical branch. In optical branch, there are four major peaks which lies around 36 meV, 43 meV, 46 meV and 50 meV energy range of phonon energies. If we refer to the plots of phonon density of states, $F(\omega)$, (in Figure 3) we find that even though the phonon density of states has a low height peak in this region (10-20 meV), the $\alpha^2F(\omega)$ has highest peak. This indicates that electron phonon coupling strength $\alpha(\omega)$ is relatively higher in this energy range (1020 meV). This indicates that electron-phonon coupling is stronger in optical branch region than in acoustical branch. The same is visible in obtained value of electron-phonon coupling constant obtained by using equation (10). The value of $\lambda$ for 1T −MoS$_2$ at ambient pressure is found to be 0.6578. This value reflects that 1T −MoS$_2$ exhibit superconductivity within weak electron-phonon coupling limit. We have also computed transition temperature ($T_c$) by employing McMillan-Allen-Dynes formula as given in equation 11. The value of Coulomb pseudopotential $\mu^*$ is taken to be 0.15. The calculated value of $T_c$ is found to be 3.5 K which is in very good agreement with recently reported experimental value of 4.0 K [25]. Taking $\mu^*$ = 0.138 gives experimental value of $T_c$. Such good agreement between calculated and experimentally measured values of Tc suggests that superconductivity in 1T −MoS$_2$ is governed by conventional electron-phonon based pairing mechanism.

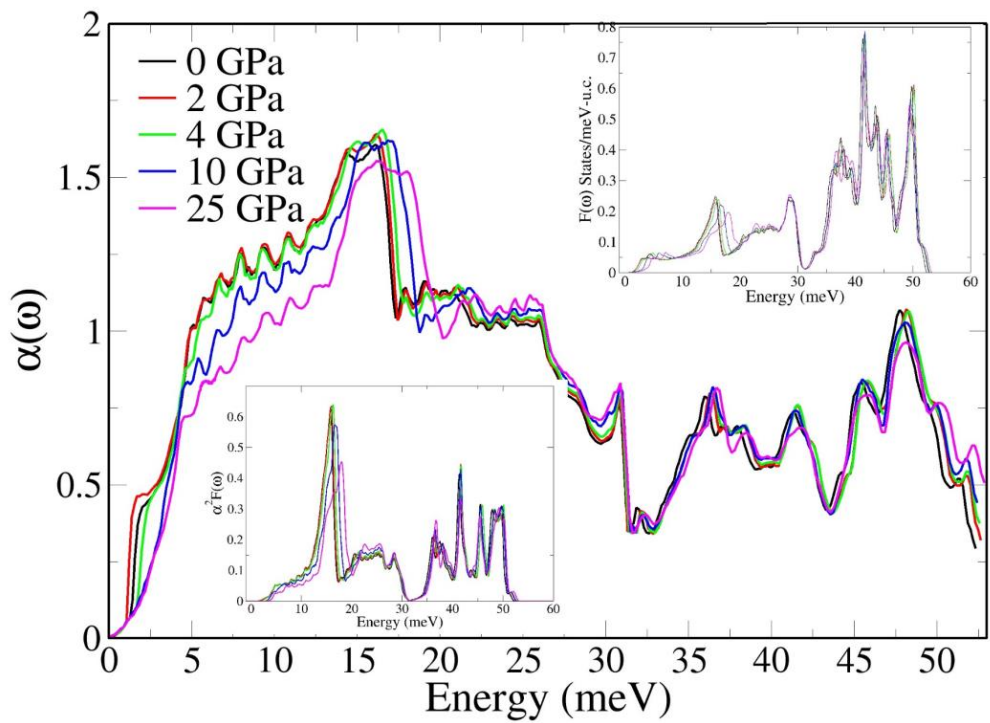

Figure 5. Electron-phonon coupling strength $\alpha(\omega)$. The upper inset shows phonon density of states $F(\omega)$ and lower inset shows Eliashberg spectral function $\alpha^2F(\omega)$ at various pressures.

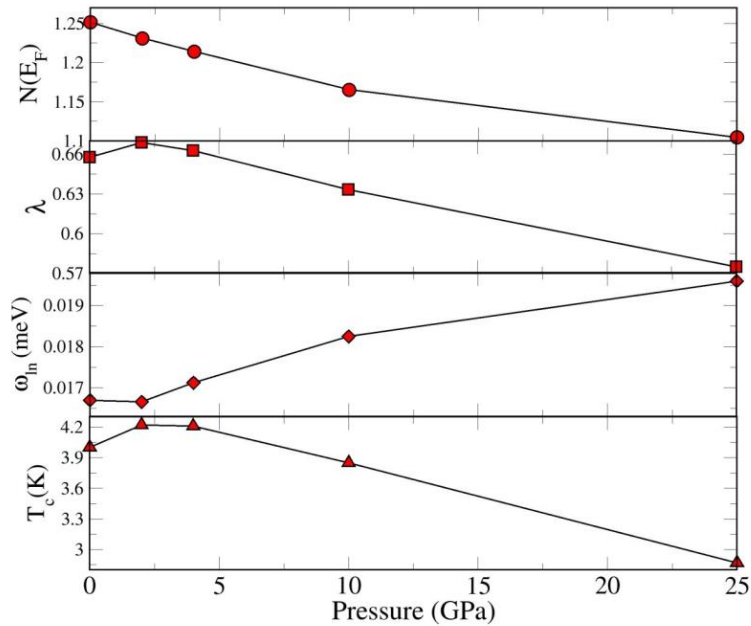

Figure 6. Pressure dependence of density of states at Fermi level $N(E_F)$, electron-phonon coupling constant ($\lambda$), average logarithmic frequency ($\omega_{ln}$) and superconducting transition temperature ($T_c$) as evaluated using McMillan-Allen-Dyne formula.

*4.3. Impact of pressure on superconductivity*

External pressure has crucial impact on superconductivity. In wake of extremely high transition temperature exhibited by electron-phonon coupled hydride superconductors, we have studied the effect of isotropic pressure on superconducting transition temperature of 1TMoS2 at 2 GPa, 4 GPa, 10 GPa and 25 GPa. In following section, we present effect of pressure on electronic, phononic and superconducting properties of 1T −MoS$_2$. The effect of pressure has been depicted in terms of change in lattice constants as obtained from equation of state fitting as depicted in Figure 1.

Variation of phonon density of states *F(ω)*, Eliashberg function *α²F(ω)* and electron phonon-coupling strength *α(ω)* with pressure is shown in Figure 5. It can be seen that with increase in pressure, there is no drastic change in any of these parameters and this is also reflected in calculated value of transition temperature. With increase in pressure, there is slight shift of the peak in electron-phonon coupling strength *α(ω)* towards higher frequency. However, with rise in pressure, Eliashberg function, *α²F(ω)*, which contributes to overall value of electron-phonon coupling constant *λ*, shows noticeable lowering in height of the peak lying in acoustical region around 15-20 meV.

Figure 6 shows variation of electronic density of states at Fermi level ($N(E_F)$), electron-phonon coupling constant (*λ*), logarithmic averaged phonon frequency ($\omega_{ln}$) and superconducting transition temperature ($T_c$) with respect to applied pressure. It can be seen that electronic density of state decreases monotonically with increase in pressure. Similar increase in ($\omega_l n$) with rise in pressure is observed. The transition temperature shows a slight increase from 4.0 K to 4.22 K at 2 GPa. There is negligible change of 0.1 K in $T_c$ when pressure changes from 2 GPa to 4 GPa. With further increase in pressure, Tc seems to decrease linearly. At 25 GPa, the Tc falls down to 2.87 K.

5. Conclusion

In this paper, we have investigated the mechanism of superconductivity in 1T − MoS₂ by employing ab-initio density functional theory. The calculated values of lattice parameters match very well with experimental values. Our calculations find that observed superconductivity in 1T −MoS₂ can be explained within conventional electron-phonon based pairing mechanism. The calculated value of transition temperature using McMillan-AllenDynes formula is in very good agreement with experimentally measured value. Our results also demonstrate that applying external pressure raises the transition temperature slightly for pressure of 2 GPa and the value does not change much from 2 GPa to 4 GPa. After 4 GPa, the transition pressure decreases linearly till 25 GPa.